\documentclass[a4paper,11pt]{article}

\usepackage{amsfonts,amsthm,amssymb}
\usepackage{latexsym}
\usepackage{graphicx}
\usepackage{axodraw}
\usepackage{color}

\newcommand{\eq}{\begin{equation}}
\newcommand{\en}{\end{equation}}
\newcommand{\eqa}{\begin{eqnarray}}
\newcommand{\ena}{\end{eqnarray}}

\def\l{\langle}
\def\r{\rangle}

\newcommand{\vrvr}[1]{|\,#1\r}
\newcommand{\lv}[1]{\l\,#1|}
\newcommand{\sss}[2]{$#1$$=$$#2$}

\newcommand{\Id}{{\mathrm{Id}}}

\newcommand{\Vcal}{{\mathcal V}}
\newcommand{\eps}{\epsilon}

\newcommand{\dx}{\!\!{\rm d}^4x\,}

\newcommand{\dk}{\!\!{\rm d}^4k\,}

\begin{document}

\thispagestyle{empty}

\begin{flushright}
hep-th/0207048: v4
\end{flushright}

\vspace*{0.8cm}

\begin{center}

{\LARGE \bf Hopf Algebraic Structures in Proving Perturbative Unitarity \\[2mm] }

\vspace{1cm}

Yong Zhang${}$
\\[0.7cm]

${}$ Institute of Theoretical Physics, Chinese Academy of Sciences\\
 P. O. Box 2735, Beijing 100080, P. R. China\\

\end{center}

\vspace{0.5cm}
\vspace{1.0cm}

\begin{center}
\parbox{12cm}{
\centerline{\small  \bf Abstract}
\small \noindent

\vspace{1.0cm}

The coproduct of a Feynman diagram is set up through identifying
the perturbative unitarity of the S-matrix with the cutting
equation from the cutting rules. On the one hand, it includes all
partitions of the vertex set of the Feynman diagram and leads to
the circling rules for the largest time equation. Its antipode is
the conjugation of the Feynman diagram. On the other hand, it is
regarded as the integration of incoming and outgoing particles
over the on-shell momentum space. This causes the cutting rules
for the cutting equation. Its antipode is an advanced function
vanishing in retarded regions. Both types of coproduct are
well-defined for a renormalized Feynman diagram since they are
compatible with the Connes--Kreimer Hopf algebra.

}

 \end{center}

 \vspace*{10mm}
 \begin{tabbing}


\end{tabbing}

 \newpage

 \section{Introduction}

  The present quantum field theory such as the standard model provides a theoretical
  description of particle physics which explains so far known high energy experiments
  \cite{peskin}. But it has to appeal to regularization and renormalization schemes
  to extract finite physical quantities from divergent Feynman diagrams. The
  renormalizability has been one basic principle for physical interesting quantum field
  theories. It is satisfied in non-abelian gauge field theories by engaging the
  spontaneously symmetry breaking mechanism \cite{thooft1}. To uncover what is behind
  successful regularization and renormalization schemes, ``some fundamental change in our
  ideas'' is needed as Dirac argued \cite{dirac}.

  Recent developments suggest that revisiting quantum field theory from Hopf algebraic points
  is one possible way out. The axioms of the Hopf algebra are listed in the appendix.
  The R-operation recipe in the BPH renormalization \cite{bogo, hepp} always
  yields local counter terms cancelling divergences of Feynman diagrams.
  It has been found to lead to the Hopf algebra of rooted trees, the Connes--Kreimer
  Hopf algebra \cite{kreimer1, kreimer2}. The Zimmermann forest formula in the BPHZ renormalization
  \cite{bogo, hepp, zim1} represents the twisted antipode of the Connes--Kreimer Hopf algebra.
  The renormalized Feynman integral has the form of convolution between the twisted antipode and
  the Feynman rules, as relates the renormalization theory to the Birkhoff decomposition and
  the Riemann-Hilbert problem \cite{kreimer2, kreimer3}. Besides the above, the Wick normal ordering
  gives a coproduct with Laplace pairs so that the Wick theorem  has a Hopf algebraic origin
  \cite{fauser1, fauser2}.

  The perturbative unitarity of the S-matrix is the perturbative realization of the unitarity of
  the S-matrix. It will be shown to have a Hopf algebraic structure. In terms of Feynman diagrams,
  it represents a diagrammatic equation, which is recognized as the cutting equation \cite{veltman}
  derived from the cutting rules \cite{cutkosky}.  The point identifies the cutting
  propagator as the integration of incoming and outgoing particles over the on-shell momentum space.
  This shows that there exists a tensor product between Feynman diagrams and also specifies a coproduct
  of the Feynman propagator. The Hopf algebra of a Feynman diagram with oriented external lines is set up
  by solving the Hopf algebraic axioms.

  The Hopf algebraic structures in the cutting rules has been introduced \cite{yong}. The cutting equation
  is an integral version of the largest time equation \cite{veltman}. The latter one is derived by the
  circling rules \cite{veltman, thooft2, vanniu} which assign an integral to a circled diagram with some
  vertexes encircled. The set of circled diagrams is obtained by partitioning the vertex set of a Feynman
  diagram. It is well known that all possible partitions of a set form a Hopf algebra. The largest time
  equation has a form of the convolution between the Feynman rules and the antipode representing the
  conjugation of a Feynman diagram.

  With the constraint of the energy conservation at every vertex, the set of circled diagrams is reduced to
  the set of admissible cut diagrams for the cutting rules. This says that the Hopf algebra in the cutting
  rules is obtained by reducing the Hopf algebra of partitions but is explained in a different way. The
  coproduct includes all admissible cuts of a Feynman diagram. The antipode is an advanced function vanishing
  in retarded region and so it does not contradict the causality principle.

  The coproducts proposed above commute with the coproduct of the Connes--Kreimer Hopf algebra so they
  are well-defined for a renormalized Feynman diagram. The paper is organized as follows. In the second
  section, through comparing the diagrammatic equation for the perturbative unitarity with the cutting
  equation, the coproduct of a Feynman diagram is specified. In the third section, the Hopf algebraic
  structures in the circling rules and the cutting rules are constructed respectively. The compatibility to
  the Connes--Kreimer Hopf algebra is considered. In the last section, the universal coproduct of a Feynman
  diagram (a set of Feynman diagrams) is discussed and further research topics are suggested. In the appendix,
  the axioms of the Hopf algebra are listed.

 \section{Coproduct in proving the perturbative unitarity}

  The equation of Feynman diagrams realizing the perturbative unitarity is found to be the
  cutting equation derived by the cutting rules \cite{veltman} in scalar field theory.

  \subsection{The perturbative unitarity of the S-matrix}

  In $\varphi^4$ model, the interaction Lagrangian density ${\cal L}_{\rm int}$ takes
  $-\frac{\lambda}{4!}\varphi^4$, $\lambda$ being coupling constant.
  Via the U-matrix approach, the S-matrix operator is given by the time-ordering
  product $T\,\exp{i\int\dx{\cal L}_{\rm int}}$  \cite{peskin}. Its hermitian $S^{\dag}$
  satisfies the unitarity equation $\label{unitarity} S\,S^{\dag}=S^{\dag}\,S=1$.
  In terms of T-matrix, $S$$=$$1+\,i\,T$, the unitarity equation has a formalism
  of matrix entries,
  \eq  \label{nppu} {\lv{f}}iT{\vrvr{i}}\,+\,{\lv{f}}(i\,T)^\dag{\vrvr{i}} =
  -\int\,{{d}^3m}\, {\lv{f}}T{\vrvr{m}}{\lv{m}}T^\dag{\vrvr{i}}, \en
  where  $\vrvr{ i}$ and ${\lv{ f}}$ respectively denote the incoming state
  $\vrvr{p_1,\,p_2,\,\cdots p_I}$ and the outgoing state $\lv{q_1,\,q_2\,\cdots q_F}$,
  the symbols $p, q$ representing on-shell external momenta, and $\vrvr{m}, \lv{m}$ denote
  intermediate on-shell states.

  In perturbative quantum field theory, the unitarity equation (\ref{nppu})
  leads to equations of Feynman diagrams. For the matrix entry $\lv{f}iT{\vrvr{i}}$, apply the Feynman
  rules, while for $\lv{f}(i\,T)^\dag\vrvr{i}$, apply the conjugation rules to conjugation
  diagrams in order to avoid treating the conjugation operator $T^\dag$. A conjugation
  diagram is a Feynman diagram except that all vertexes are encircled. The conjugation
  rules assign the factor of $\int\,\dx\,(i\lambda)$ to the vertex $x$
  and the conjugation propagator $\Delta_F^\ast$, the complex conjugation of
  $\Delta_F$, to each internal line.

  With connected Feynman diagrams, the unitarity equation
  (\ref{nppu}) shows in a diagrammatic form
 \begin{center}
 \begin{picture}(300,100)(0,0)
 \GBox(20,50)(40,70){.5}
 \ArrowLine(10,90)(20,70)
 \ArrowLine(10,30)(20,50)\Text(10,60)[]{$i$}
 \ArrowLine(40,70)(60,90) \ArrowLine(40,50)(60,30)
 \DashLine(70,90)(70,30){5} \Text(50,60)[]{$f$}
 \Text(80,60)[]{$+$} \DashLine(90,90)(90,30){5}
\GBox(110,50)(130,70){.5}
 \ArrowLine(100,90)(110,70)
 \ArrowLine(100,30)(110,50)\Text(100,60)[]{$i$}
 \ArrowLine(130,70)(150,90) \ArrowLine(130,50)(150,30)
 \Text(160,60)[]{$=$} \Text(190,60)[]{$-\int d^3 m$}
\GBox(230,50)(240,70){.5}
 \ArrowLine(220,90)(230,70)
 \ArrowLine(220,30)(230,50)\ArrowLine(240,70)(255,85)
 \ArrowLine(240,50)(255,35)\DashLine(260,90)(260,30){5}
 \Text(220,60)[]{$i$}\Text(250,60)[]{$m$}
\GBox(280,50)(290,70){.5}
 \ArrowLine(265,85)(280,70)
 \ArrowLine(265,35)(280,50)\ArrowLine(290,70)(310,90)
 \ArrowLine(290,50)(310,30)
 \Text(300,60)[]{$f$}\Text(270,60)[]{$m$}
 \end{picture}
 \vspace{-.5cm}
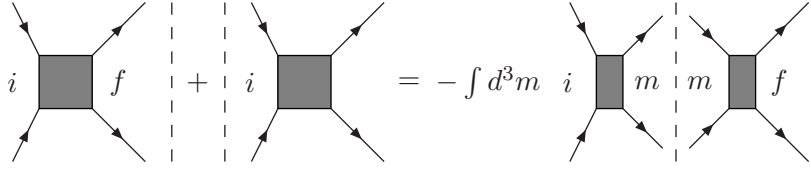
  {\\ Figure 1. The unitarity equation of Feynman diagrams.}
  \end{center}
 As a convention, incoming external lines are on the left hand side of the diagram and
 outgoing external lines on the right hand side so that the positive energy always flows from the
 left to the right. The dashed line distinguishes the Feynman diagram on its left hand side from
 the conjugation diagram on its right hand side.

 \subsection{Coproduct in proving the perturbative unitarity}

 The perturbative unitarity is proved by identifying the unitarity equation with the cutting equation
 \cite{veltman, thooft2, vanniu}. For scalar field theories, the cutting
 equation \cite{vanniu} has a diagrammatic form,
 \begin{center}
 \begin{picture}(300,100)(0,0)
 \GBox(20,50)(40,70){.5}
 \ArrowLine(10,90)(20,70)
 \ArrowLine(10,30)(20,50)    \Text(10,60)[]{$i$}
 \ArrowLine(40,70)(60,90)    \ArrowLine(40,50)(60,30)
 \DashLine(70,90)(70,30){5}  \Text(50,60)[]{$f$}
 \Text(80,60)[]{$+$} \DashLine(90,90)(90,30){5}
 \GBox(110,50)(130,70){.5}
 \ArrowLine(100,90)(110,70)
 \ArrowLine(100,30)(110,50)\Text(100,60)[]{$i$}
 \ArrowLine(130,70)(150,90) \ArrowLine(130,50)(150,30)
 \Text(140,60)[]{$f$}
 \Text(160,60)[]{$=$}
 \Text(190,60)[]{$-\Sigma_{\textrm{a.c.}}$}
 \GBox(230,50)(240,70){.5}
 \ArrowLine(220,90)(230,70)
 \ArrowLine(220,30)(230,50) \DashLine(260,90)(260,30){5}
 \Text(220,60)[]{$i$} \CArc(260,50)(28.28,45,135)
 \GBox(280,50)(290,70){.5}
 \ArrowLine(290,70)(310,90)\CArc(260,70)(28.28,225,315)
 \ArrowLine(290,50)(310,30) \Text(300,60)[]{$f$}
 \Text(260,60)[]{a.c.}
 \end{picture}
 \vspace{-.5cm}
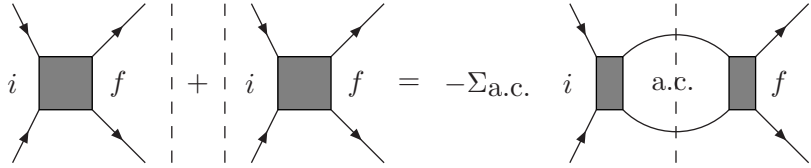
  {\\ Figure 2. The cutting equation in Feynman diagrams.}
  \end{center}
 Here the dashed line represents the cut line through a diagram and the symbol ``a.c." means
 admissible cutting. The cutting rules and the cutting equation will be introduced in the
 next section.

 The right part of Fig.2 has to be recognized to be the right part of Fig.1. Hence the cutting
 propagator $\Delta_+$ is required to be decomposed into an integration over the phase space of an incoming
 particle and outgoing particle,
  \eq
  \label{coproduct}
  \Delta_{+}(x-y)=\int\,{\rm d}^3k\, {\frac
  {e^{-ik\,y}}{\sqrt{(2\pi)^3 2\omega_k}}}\, {\frac
  {e^{ik\,x}}{\sqrt{(2\pi)^3 2\omega_k}}} \en
 where $\omega_k$$=$$\sqrt{ {k}^2+m^2}$ and $m$ denotes mass of
 particle. The positive (negative) cutting propagators $\Delta_+$ ($\Delta_-$) takes the form
  \eq
  \Delta_{+}(x-y)=\Delta_{-}(y-x)=\int\,{\frac {\dk}
  {(2\pi)^4}}\,\, \theta(k_0)\, 2\pi\,\delta(k^2+m^2) \,e^{i\,k(x-y)},
  \en
 which are combined into the Feynman propagator $\Delta_F$,
  \eq
  \Delta_F(x-y)=\theta(x^0-y^0)\,\Delta_{+}(x-y)+ \theta(y^0-x^0)\,\Delta_{-}(x-y),
  \en
 $\theta$ function being the normal step function.  The four propagators: $\Delta_F$, $\Delta_F^{\ast}$,
 $\Delta_{+}$ and $\Delta_{-}$ satisfy
 \eq \label{simplest}
 (\Delta_F+\Delta^\ast_F-\Delta_{+} -\Delta_{-})(x-y)=0 \en  which is an  example of the largest time
  equation \cite{veltman, thooft2, vanniu}.

  Through comparing Fig.1 with Fig 2,  we observe a tensor product structure in the proof for the perturbative
  unitarity.  The equation (\ref{coproduct}) is regarded as a multiplication $m$
  representing the integration over the tensor product of Feynman
  diagrams,
 \begin{center}
  \begin{picture}(100,30)(0,0)
  \Text(0,20)[]{$m($}
  \Vertex(20,20){1.5} \Line(10,20)(20,20)
  \Text(30,20)[]{$\otimes$}
  \Vertex(40,20){1.5} \Line(40,20)(50,20)
   \Text(60,20)[]{$ )$}
 \Text(40,18)[lt]{$x$} \Text(20,18)[rt]{$y$}
  \Text(70,20)[]{$=$}  
  \Line(83,20)(100,20)
  \Vertex(100,20){1.5} \CArc(81.5,20)(1.5,0,360)
  \Text(81.5,18)[lt]{$x$} \Text(100,18)[rt]{$y$}        
   \end{picture}
   \vspace{-.3cm}
  {\\ Figure 3. Tensor product in proving the perturbative unitarity.}
  \end{center}
  In terms of such tensor products,  we construct a coproduct of the Feynman propagator
  $\Delta_F(x-y)$,
 \begin{center}
  \begin{picture}(280,20)(0,0)
  \Text(-5,10)[]{$\Delta ($}        
  \Line(5,10)(25,10)
  \Vertex(5,10){1.5}\Vertex(25,10){1.5}
  \Text(5,8)[lt]{$x$} \Text(25,8)[rt]{$y$}
  \Text(30,10)[]{$)$}              
  \Text(40,10)[]{$=$}
  \Vertex(50,10){1.5}\Vertex(70,10){1.5}\Line(50,10)(70,10)
  \Text(50,8)[lt]{$x$} \Text(70,8)[rt]{$y$}
  \Text(80,10)[]{$\otimes$}
  \Text(90,10)[]{$e$}                  
  \Text(100,10)[]{$+$}
  \Text(110,10)[]{$e$}  \Text(120,10)[]{$\otimes$}
  \Vertex(130,10){1.5}  \Vertex(150,10){1.5}\Line(130,10)(150,10)
  \Text(130,8)[lt]{$x$} \Text(150,8)[rt]{$y$}     
  \Text(160,10)[]{$+$}
  \Vertex(170,10){1.5}\Line(170,10)(180,10)
   \Text(170,8)[lt]{$x$}
  \Text(190,10)[]{$\otimes$}
  \Vertex(210,10){1.5}\Line(200,10)(210,10)
   \Text(210,8)[rt]{$y$}                     
  \Text(220,10)[]{$+$}
  \Vertex(240,10){1.5}\Line(230,10)(240,10)
  \Text(240,8)[rt]{$y$}
  \Text(250,10)[]{$\otimes$}
  \Vertex(260,10){1.5}\Line(260,10)(270,10)
   \Text(260,8)[lt]{$x$}                      
  \end{picture}
   {\vspace{.3cm}\\ Figure 4. The coproduct of the Feynman propagator. }
  \end{center}
  The symbol $e$ denotes the unit of Hopf algebra or the empty set $\varnothing$. With
  the character $\Phi$ representing the
  Feynman rules and the character $\Phi_c$ representing the conjugation rules, the largest time equation
  (\ref{simplest}) has an algebraic formulation $m(\Phi\otimes\Phi_c)\Delta(\Delta_F)=0$, namely the
  convolution $\Phi\star\Phi_c(\Delta_F)$ between two characters $\Phi$ and $\Phi_c$.

 \section{Hopf algebraic structures in the cutting rules}

 With the above coproduct of the Feynman propagator, two types of Hopf algebras are set up. The
 first one reflects the circling rules  \cite{veltman} for the largest time
 equation. The second one is related to the cutting rules for the cutting equation \cite{veltman}.
 Both survive renormalization at least the dimensional regularization and the minimal subtraction.

 \subsection{The Hopf algebra in the circling rules}

 Motivated by the simplest largest time equation (\ref{simplest}) of the Feynman propagator, we
 are going to construct the largest time equation for an arbitrary $N$-vertex connected Feynman
 diagram $\Gamma$ in terms of  $\Delta_{\pm}$, $\Delta_F$
 and  $\Delta_F^\ast$. With the observation that the coproduct in  Fig.4 is regarded
  as a complete partition of the vertex set $\{x, y\}$, the circling rules \cite{veltman} is devised and
 applied to a set of circled diagrams.

 The Feynman diagram $\Gamma$ is specified by both the set $\Vcal_N$ of its
 vertexes, $\Vcal_N=\{x_1,x_2,\cdots x_N \}$ and the set of all lines connecting vertexes $x_i$. It corresponds
 to $2^N$ circled diagrams with some vertexes encircled. The case of no vertexes encircled is $\Gamma$
 itself and the case of all vertexes encircled is its conjugation diagram $\Gamma^\ast$.

 The circling rules map a circled diagram to an integral formalism. For a vertex $x$, assign
 the factor of $\int\,d^4 x\,(i\lambda)$; for an internal line connecting two circled vertexes, assign
 the conjugation propagator $\Delta_F^\ast$; for an internal line connecting a circled vertex $x_i$ to
 a uncircled vertex $x_j$, assign the positive cutting propagator $\Delta_{+}(x_i-x_j)$, namely for
 an internal line connecting a uncircled vertex $x_i$ to a circled vertex $x_j$, assigns the negative
 cutting propagator $\Delta_{-}(x_i-x_j)$; for other ingredients of a circled diagram, apply the Feynman
 rules.  As an example of  applying the circling rules, a two-loop four-point Feynman
 diagram $\Gamma_{\mbox{\tiny Fig.5}}$ in $-\frac {\lambda} {4!} \varphi^4$ model has its
 largest time equation
\begin{center}
  \begin{picture}(320,40)(0,0)
  \Vertex(0,20){1.5} \Vertex(20,30){1.5}  \Vertex(20,10){1.5}
   \Line(1.5,20)(20,28.5) \Line(1.5,20)(20,11.5) \Oval(20,20)(8.5,3)(0)
   \Text(30,20)[]{$+$}
    \BCirc(40,20){1.5} \BCirc(60,30){1.5}
    \BCirc(60,10){1.5}
   \Line(41.5,20)(60,28.5) \Line(41.5,20)(60,11.5) \Oval(60,20)(8.5,3)(0)
   \Text(70,20)[]{$+$}
    \BCirc(80,20){1.5} \Vertex(100,30){1.5}   \Vertex(100,10){1.5}
   \Line(81.5,20)(100,28.5) \Line(81.5,20)(100,11.5) \Oval(100,20)(8.5,3)(0)
   \Text(110,20)[]{$+$}
   \Vertex(120,20){1.5} \BCirc(140,30){1.5}  \Vertex(140,10){1.5}
   \Line(121.5,20)(140,28.5) \Line(121.5,20)(140,11.5) \Oval(140,20)(8.5,3)(0)
   \Text(150,20)[]{$+$}
   \Vertex(160,20){1.5} \Vertex(180,30){1.5}  \BCirc(180,10){1.5}
   \Line(161.5,20)(180,28.5) \Line(161.5,20)(180,11.5) \Oval(180,20)(8.5,3)(0)
   \Text(190,20)[]{$+$}
   \BCirc(200,20){1.5} \BCirc(220,30){1.5}  \Vertex(220,10){1.5}
   \Line(201.5,20)(220,28.5) \Line(201.5,20)(220,11.5) \Oval(220,20)(8.5,3)(0)
   \Text(230,20)[]{$+$}
   \BCirc(240,20){1.5} \BCirc(260,30){1.5}  \Vertex(260,10){1.5}
   \Line(241.5,20)(260,28.5) \Line(241.5,20)(260,11.5) \Oval(260,20)(8.5,3)(0)
   \Text(270,20)[]{$+$}
   \Vertex(280,20){1.5} \BCirc(300,30){1.5}  \BCirc(300,10){1.5}
   \Line(281.5,20)(300,28.5) \Line(281.5,20)(300,11.5) \Oval(300,20)(8.5,3)(0)
   \Text(310,20)[]{$=$}
   \Text(320,20)[]{$0$}
   \end{picture}
{ \\Figure 5. An example for the largest time equation.}
  \end{center}

 In the following, a Hopf algebra in the circling rules is set up.
 $H$ is a set generated by all connected Feynman
 diagrams with oriented external lines. $F$ is a field representing the complex number $\mathbb{C}$ with
 unit $1$. The addition $+$ is defined by the linear combination
 $a\,\Gamma_1\,+\,b\,\Gamma_2\in H, a, b\in\mathbb{C}, \Gamma_1, \Gamma_2\in H$
 from which the triple $(H,+;F)$ is a vector space. The multiplication  $m$
 of two Feynman diagrams $\Gamma_1$ and $\Gamma_2$ is specified by their disjoint union
 $\Gamma_1\Gamma_2=m(\Gamma_1\otimes\Gamma_2):=\Gamma_1\cup\Gamma_2 $. $\eta$ is the unit map
 \sss{\eta(1)}{e} and specifies the empty set
 $\varnothing$ as the unit $e$.  With these definitions, the associativity axiom $(1)$ and
 the unit axiom $(2)$ are satisfied.

 For a Feynman diagram $\Gamma$, the coproduct $\Delta$ involves its subdiagrams and reduced subdiagrams.
 The subdiagram  $\gamma(\Vcal_c)$ is specified by both the subset $\Vcal_c$ of $\Vcal_N$ and the set of
 all lines connecting vertexes in $\Vcal_c$. The reduced subdiagram $\Gamma/\gamma$ is obtained by
 cutting the subdiagram $\gamma$ out of $\Gamma$. Hence the coproduct  $ \Delta$ is
 defined by $\Delta(\Gamma)=\sum_{P}\,\gamma(\Vcal_c)\otimes\,\Gamma/\gamma$, where $P$ denotes all
 possible partitions of the vertex set $\Vcal_N$. It expands as
 \eq
 \Delta(\Gamma)=\, \Gamma\otimes e+\, e\otimes \Gamma +
 \sum_{1\leq c< N}\,\gamma(\Vcal_c)\otimes\,\Gamma/\gamma,
 \en
 where the summation is over all subdiagrams except $\varnothing$ and $\Gamma$.
 External lines of $\gamma(\Vcal_c)$ from cut internal lines are outgoing, while external lines of
 $\Gamma/\gamma$ from cut internal lines are incoming. As an example, the coproduct of an one-loop four-point
 Feynman diagram $\Gamma_{\mbox{\tiny Fig.6}}$ in the $-\frac {\lambda} {4!}\varphi^4$ model has the
 diagrammatic representation
\begin{center}
  \begin{picture}(230,40)(0,0)
 \Text(-5,30)[]{$\Delta ($}
 \Vertex(10,30){1.5} \Vertex(30,30){1.5}
 \Text(10,26)[rt]{$x$} \Text(30,26)[lt]{$y$}
 \Oval(20,30)(5,10)(0)  \Text(40,30)[]{$)$}  
 \Text(50,30)[]{$=$}\Vertex(60,30){1.5} \Vertex(80,30){1.5}
 \Text(60,26)[rt]{$x$} \Text(80,26)[lt]{$y$}
 \Oval(70,30)(5,10)(0)
 \Text(90,30)[]{$\otimes$} \Text(100,30)[]{$e$}    
 \Text(110,30)[]{$+$}\Text(120,30)[]{$e$}\Text(130,30)[]{$\otimes$}
 \Vertex(140,30){1.5} \Vertex(160,30){1.5}
 \Text(140,26)[rt]{$x$} \Text(160,26)[lt]{$y$}
 \Oval(150,30)(5,10)(0)      
   \Text(170,30)[]{$+$}
   \Vertex(180,30){1.5}
   \Line(180,30)(190,40)  \Line(180,30)(190,20) \Text(180,26)[tr]{$x$}
   \Text(200,30)[]{$\otimes$} \Vertex(220,30){1.5}
  \Line(220,30)(210,40) \Line(220,30)(210,20)
   \Text(220,26)[tl]{$y$}  
  \Text(230,30)[]{$+$}
 \Vertex(250,30){1.5}
  \Line(240,40)(250,30)  \Line(240,20)(250,30) \Text(250,26)[tl]{$y$}

  \Text(260,30)[]{$\otimes$} \Vertex(270,30){1.5}
  \Text(270,26)[tr]{$x$} \Line(280,20)(270,30) \Line(280,40)(270,30)
  \end{picture}
  \vspace{-.5cm}
{\\ Figure 6. An example for the coproduct in the circling rules.}
  \end{center}

 The counit $\eps$ is chosen to satisfy $\eps(e)=1; \eps(\Gamma)=0,
 \mbox{if}~ \Gamma\neq e$. With our choices of the coproduct
 and counit, the coassociativity axiom (5) and the counit axiom (6)
 are ensured. Obviously, the process of dividing the vertex
 set and further dividing the vertex set of the subdiagram is
 equivalent to that of dividing the vertex set and further dividing
 the vertex set of the reduced subdiagram. At last, solving the equation
 representing the antipode axiom $(8)$, the antipode $S$ is obtained to be
 $S(\Gamma)=-\Gamma-\sum_{1\leq c<N}\,S(\gamma(\Vcal_c))\,\Gamma/\gamma$ with
 $S(e)=e$.

 In terms of the Feynman rules $\Phi$ and the conjugation rules $\Phi_c$, the
 largest time equation of  $\Gamma$ has a form of convolution
 $\Phi\star \Phi_c (\Gamma)=\Phi_c\star\Phi (\Gamma)=0$. As an example, the
 coproduct of the Feynman diagram $\Gamma_{\mbox{\tiny Fig.6}}$ leads to the
 largest time equation
 \eq
 \Phi\star\Phi_c (\Gamma_{\mbox{\tiny Fig.6}})
 =\Delta_F^2+\Delta_F^{\ast 2}-\Delta_+^2-\Delta_-^2=0.
 \en

 We go further to calculate the antipode $S(\Delta_F)$ of the Feynman
 propagator,
\begin{center}
\begin{picture}(170,40)(0,0)
  \Text(0,30)[]{$S($}
  \Vertex(10,30){1.5}    \Vertex(30,30){1.5}
  \Line(10,30)(30,30)    \Text(40,30)[]{$ )$}
 \Text(10,28)[lt]{$x$} \Text(30,28)[rt]{$y$}  
  \Text(50,30)[]{$=$}    \Text(60,30)[]{$-$}
  \Line(70,30)(90,30) \Vertex(70,30){1.5}
   \Vertex(90,30){1.5}
  \Text(70,28)[lt]{$x$} \Text(90,28)[rt]{$y$}      
   \Text(100,30)[]{$+$}
    \Vertex(110,30){1.5}   \CArc(128.5,30)(1.5,0,360)
     \Line(110,30)(127,30)
    \Text(110,28)[lt]{$x$} \Text(128.5,28)[rt]{$y$}       
    \Text(140,30)[]{$+$}
    \CArc(151.5,30)(1.5,0,360)   \Vertex(170,30){1.5}
     \Line(153,30)(170,30)
 \Text(151.5,28)[lt]{$x$} \Text(170,28)[rt]{$y$}
   \end{picture}
   \vspace{-.5cm}
  {\\  Figure 7. The antipode of the Feynman propagator.}
  \end{center}
 With the simplest largest time equation (\ref{simplest}), $S(\Delta_F)$ is the conjugation propagator
 $\Delta_F^\ast$. This suggests that the antipode $S(\Phi(\Gamma))$ of the Feynman diagram $\Gamma$ is
 obtained by applying the conjugation rules to its conjugation diagram $\Gamma^\ast$, namely,
 $S(\Phi(\Gamma))=\Phi_c\,(\Gamma^\ast)$. Therefore the largest time equation has a
typical Hopf algebraic
 form $S\star\Phi(\Gamma)=\Phi\star S(\Gamma)=0$.

\subsection{The Hopf algebra in the cutting rules}

Circled diagrams with a vertex connecting other vertexes by the
positive (negative) cutting propagator have vanishing Feynman
integrals. At such a vertex, the conservation of energy is
violated since $\Delta_+$ ($\Delta_-$) only admits the positive
(negative) energy flow. An admissible cut diagram is  suitable to
represent a non-vanishing circled diagram. For a connected Feynman
diagram $\Gamma$, make a cut line  through its internal or
external lines to separate it into two parts. The left part
denotes the admissible cut diagram $\gamma$ with at least one
incoming external line, while the right part denotes the
admissible reduced cut diagram $\Gamma/\gamma$ with at least one
outgoing external line. The cases of only cutting incoming or
outgoing lines are included.

The cutting rules \cite{veltman, thooft2, vanniu} assign a Feynman
integral to a cut diagram. Apply the Feynman rules to $\gamma$ and
the conjugation rules to $\Gamma/\gamma$; assign $\Delta_+(x-y)$
to a cut internal line with the vertex $x$ in $\gamma$ and the
vertex $y$ in $\Gamma/\gamma$. As an example, the cutting equation
of the Feynman diagram $\Gamma_{\mbox{\tiny Fig.5}}$ at $x_1^0 <
x_2^0, x_3^0$  has a diagrammatic form,
\begin{center}
 \begin{picture}(240,50)(0,0)
 \Vertex(0,20){1.5} \Vertex(20,30){1.5}  \Vertex(20,10){1.5}
 \Text(0,15)[t]{$x_1$} \Text(20,5)[t]{$x_3$} \Text(20,35)[b]{$x_2$}
 \Line(1.5,20)(20,28.5) \Line(1.5,20)(20,11.5) \Oval(20,20)(8.5,3)(0)
 \DashLine(30,30)(30,10){3}
 \Text(40,20)[]{$+$}
 \DashLine(50,30)(50,10){3}
 \BCirc(60,20){1.5} \BCirc(80,30){1.5}  \BCirc(80,10){1.5}
 \Text(60,15)[t]{$x_1$} \Text(80,5)[t]{$x_3$} \Text(80,35)[b]{$x_2$}
 \Line(61.5,20)(80,28.5) \Line(61.5,20)(80,11.5) \Oval(80,20)(8.5,3)(0)
 \Text(90,20)[]{$+$}
 \DashCArc(120,30)(10,180,330){3}
 \Vertex(100,20){1.5} \BCirc(120,30){1.5}  \Vertex(120,10){1.5}
 \Text(100,15)[t]{$x_1$} \Text(120,5)[t]{$x_3$} \Text(120,35)[b]{$x_2$}
 \Line(101.5,20)(120,28.5) \Line(101.5,20)(120,11.5) \Oval(120,20)(8.5,3)(0)
 \Text(140,20)[]{$+$}
  \Vertex(150,20){1.5}    \Vertex(170,30){1.5}  \BCirc(170,10){1.5}
  \Text(150,15)[t]{$x_1$} \Text(170,5)[t]{$x_3$} \Text(170,35)[b]{$x_2$}
  \Line(151.5,20)(170,28.5)  \Line(151.5,20)(170,11.5) \Oval(170,20)(8.5,3)(0)
  \DashCArc(170,10)(10,0,180){3}
  \Text(190,20)[]{$+$}
  \Vertex(200,20){1.5} \BCirc(220,30){1.5}  \BCirc(220,10){1.5}
  \Text(200,15)[t]{$x_1$} \Text(220,5)[t]{$x_3$} \Text(220,35)[b]{$x_2$}
  \Line(201.5,20)(220,28.5) \Line(201.5,20)(220,11.5) \Oval(220,20)(8.5,3)(0)
  \DashLine(212,30)(212,10){3}
  \Text(230,20)[]{$=$}
  \Text(240,20)[]{$0$}
  \end{picture}

   \vspace{.3cm}
  { Figure 8. An example for the cutting equation at $x_1^0 < x_2^0, x_3^0$.}
  \end{center}

The Hopf algebra in the cutting rules is regarded as a reduced version of the Hopf
algebra in the circling rules by imposing the conservation of energy at each vertex
of a Feynman diagram.  The set of admissible cut diagrams of the Feynman diagram $\Gamma$
specifies the coproduct
 \eq \Delta(\Gamma)=
\Gamma\,\otimes e+ e\otimes \Gamma\,+\sum_{\mbox{a.c.}}\,\gamma\otimes\,\Gamma/\gamma,
 \en
 where the symbol ``a.c.'' means admissible cutting and the summation excludes cases of
 only cutting external lines. As an example, the coproduct of the Feynman diagram
 $\Gamma_{\mbox{\tiny Fig.6}}$ at $y^0 < x^0 $  has a diagrammatic form
\begin{center}
  \begin{picture}(170,40)(0,0)
 \Text(-5,30)[]{$\Delta ($}
 \Vertex(10,30){1.5} \Vertex(30,30){1.5}
 \Text(10,26)[rt]{$x$} \Text(30,26)[lt]{$y$}
 \Oval(20,30)(5,10)(0)  \Text(40,30)[]{$)$}  
 \Text(50,30)[]{$=$}\Vertex(60,30){1.5} \Vertex(80,30){1.5}
 \Text(60,26)[rt]{$x$} \Text(80,26)[lt]{$y$}
 \Oval(70,30)(5,10)(0)
 \Text(90,30)[]{$\otimes$} \Text(100,30)[]{$e$}    
 \Text(110,30)[]{$+$}\Text(120,30)[]{$e$}\Text(130,30)[]{$\otimes$}
 \Vertex(140,30){1.5} \Vertex(160,30){1.5}
 \Text(140,26)[rt]{$x$} \Text(160,26)[lt]{$y$}
 \Oval(150,30)(5,10)(0)      
  \Text(170,30)[]{$+$}
 \Vertex(190,30){1.5}
  \Line(180,40)(190,30)  \Line(180,20)(190,30) \Text(190,26)[tl]{$y$}
  \Text(200,30)[]{$\otimes$} \Vertex(210,30){1.5}
  \Text(210,26)[tr]{$x$} \Line(220,20)(210,30) \Line(220,40)(210,30)
  \end{picture}
  \vspace{-.5cm}
  {\\ Figure 9.  An example for the coproduct in the cutting rules.}
  \end{center}

In terms of the characters $\Phi$ and $\Phi_c$, the cutting
equation has an coalgebraic form
$m(\Phi\otimes\Phi_c)\Delta(\Gamma)=0$. Such a coproduct satisfies
the coassociativity axiom $(3)$. Cutting $\Gamma$ twice leads to
the set of all possible admissible cut diagrams
$\gamma_1\otimes\gamma_2\otimes\gamma_3$ which is irrelevant to
the order of two cuttings. For example, the Feynman diagram
$\Gamma_{\mbox{\tiny Fig.5}}$ has a coassociative product, as is
verified by
 \begin{center}
 \begin{picture}(190,40)(0,0)
 \Text(0,20)[]{$(\Delta \otimes {\rm Id} ) \Delta$}
 \Text(30,20)[]{$($}
  \Vertex(40,20){1.5} \Vertex(60,30){1.5}  \Vertex(60,10){1.5}
 \Text(40,15)[t]{$x_1$} \Text(60,5)[t]{$x_3$} \Text(60,35)[b]{$x_2$}
 \Line(41.5,20)(60,28.5) \Line(41.5,20)(60,11.5) \Oval(60,20)(8.5,3)(0)
  \Text(70,20)[]{$)$} \Text(80,20)[]{$=$}
  \Text(120,20)[]{$({\rm Id}\otimes \Delta) \Delta$ }
  \Text(150,20)[]{$($}
   \Vertex(160,20){1.5} \Vertex(180,30){1.5}  \Vertex(180,10){1.5}
 \Text(160,15)[t]{$x_1$} \Text(180,5)[t]{$x_3$} \Text(180,35)[b]{$x_2$}
 \Line(161.5,20)(180,28.5) \Line(161.5,20)(180,11.5) \Oval(180,20)(8.5,3)(0)
  \Text(190,20)[]{$)$}
  \end{picture}

{ Figure 10. An example for the coassociative coproduct in the
cutting rules.}
  \end{center}

As the above, a coalgebraic structure has been set up. The
bialgebraic axiom $(5)$ and $(6)$ have to be verified.  To
understand the axiom $(5)$,  study the Feynman diagram
$\Gamma_{\mbox{\tiny Fig.5}}$ as an example
\begin{center}
 \begin{picture}(130,40)(0,0)
 \Text(0,20)[]{$\Delta\,($}
  \Vertex(15,20){1.5} \Line(16.5,20)(25,30) \Line(16.5,20)(25,10)
  \Vertex(30,20){1.} \Line(35,28.5)(45,28.5)\Line(35,11.5)(45,11.5)
  \Vertex(45,30){1.5} \Vertex(45,10){1.5}
  \Oval(45,20)(8.5,3)(0)
  \Text(15,15)[t]{$x_1$} \Text(45,35)[b]{$x_2$} \Text(45,5)[t]{$x_3$}
  \Text(55,20)[]{$)$}
   \Text(65,20)[]{$=$} \Text(80,20)[]{$\Delta\,($}
   \Vertex(95,20){1.5}
   \Line(96.5,20)(105,30) \Line(96.5,20)(105,10)
   \Text(110,20)[]{$)$}
   \Vertex(115,20){1.} \Text(130,20)[]{$\Delta\,($}
   \Line(140,28.5)(150,28.5)\Line(140,11.5)(150,11.5)
   \Vertex(150,30){1.5} \Vertex(150,10){1.5}
   \Oval(150,20)(8.5,3)(0)
   \Text(160,20)[]{$)$}
  \Text(95,15)[t]{$x_1$} \Text(150,35)[b]{$x_2$} \Text(150,5)[t]{$x_3$}
  \end{picture}

{ Figure 11. An example for the bialgebraic axiom.}
  \end{center}
In the case that the multiplication between two Feynman diagrams
is the disjoint union, the diagrammatic equation in Fig. 11 is
regarded as a definition of the coproduct. In the other case that
the multiplication represents the integration over the phase space
of incoming and outgoing particles, it survives Feynman integrals.
Considering Feynman integrands, its right hand side has more terms
than its left hand side.  With the choices of the coproduct and
counit, the axiom $(6)$ is easily checked to be satisfied.

Solving the antipode axiom $(7)$,  the antipode of the Feynman
diagram $\Gamma$ in the cutting rules has the form $
S(\Gamma)=-\Gamma- \sum_{\mbox{a.c.}}\,S(\gamma)\,\Gamma/\gamma $
with $S(e)=e$. The antipode of the Feynman propagator $\Delta_F$
takes the  form
\begin{center}
  \begin{picture}(130,30)(0,0)
  \Text(0,30)[]{$S($}
  \Vertex(10,30){1.5}    \Vertex(30,30){1.5}
  \Line(10,30)(30,30)    \Text(40,30)[]{$ )$}
 \Text(10,28)[lt]{$x$} \Text(30,28)[rt]{$y$}         
  \Text(50,30)[]{$=$}    \Text(60,30)[]{$-$}
  \Line(70,30)(90,30) \Vertex(70,30){1.5}
   \Vertex(90,30){1.5}
  \Text(70,28)[lt]{$x$} \Text(90,28)[rt]{$y$}        
    \Text(100,30)[]{$+$}
    \CArc(111.5,30)(1.5,0,360)   \Vertex(130,30){1.5}
    \Line(113,30)(130,30)
 \Text(111.5,28)[lt]{$x$} \Text(130,28)[rt]{$y$}
   \end{picture}
   \vspace{-.5cm}
  { \\ Figure 12. The antipode of the Feynman propagator at $x^0 > y^0$.}
  \end{center}
Its formalism of Feynman integrands is given by \eq
S(\Delta_F(x))=\theta(-x^0)(\Delta_+(x)-\Delta_-(x))
 \en
with $y=0$, which vanishes in the retarded region. As a
generalization, the antipode in the cutting rules is an advanced
function. Calculating the antipode of $\Gamma_{\mbox{\tiny
Fig.5}}$ at $x_1^0 < x_2^0 < x_3^0$ to obtain \eqa
S(\Gamma_{\mbox{\tiny Fig.5}}) &=&
(\Delta_+(x_{21})\Delta_+(x_{31})-\Delta_F(x_{21})\Delta_F(x_{31}))
\Delta_F^2(x_{23}) \nonumber\\
&+&(\Delta_F(x_{21})-\Delta_+(x_{21}) )
\Delta_+(x_{31})\Delta_-^2(x_{23}) \ena where $x_{21}=x_2-x_1$,
$x_{31}=x_3-x_1$ and $x_{23}=x_2-x_3$. It vanishes at the retarded
region $x_1^0 < x_2^0 < x_3^0$. Hence the Hopf algebraic structure
in the cutting rules cooperates the perturbative unitarity of the
S-matrix with its causality.

 \subsection{The Hopf algebraic structures under renormalization}

 For a divergent Feynman diagram $\Gamma$, the renormalized largest time equation and the renormalized cutting
 equation can be set up under the dimensional regularization and the minimal subtraction. They also have
 Hopf algebraic representations similar to the preceding constructions but involving the Connes--Kreimer Hopf
 algebra \cite{kreimer1, kreimer2, kreimer3} denoting the BPH renormalization \cite{bogo,hepp}.

 The Connes--Kreimer Hopf algebra is defined in the space of 1PI Feynman diagrams. The product of two Feynman
 diagrams is their disjoint union. The coproduct of a divergent Feynman diagram $\Gamma$ expresses all possible
 disjoint unions of its divergent subdiagrams
 \eq
 \Delta_{CK}(\Gamma)=\Gamma\otimes\,e+\, e\otimes \Gamma+\sum\,\gamma\otimes\Gamma/\gamma.
 \en
 where $\gamma\subset \Gamma$. The antipode is obtained by solving the antipode axiom $(7)$.
 The global counter term for $\Gamma$ is given by the twisted antipode
 \eq
 S_R(\Gamma)=-{\cal R}(\Phi(\Gamma))-{\cal R} \sum\,S_R(\gamma)\,\Phi(\Gamma/\gamma)
 \en
 where $\cal R$ denotes the minimal subtraction in the dimensional renormalization. The convolution
 $S_R\star\Phi(\Gamma)$ leads a renormalized Feynman integral corresponding
 to the bare one $\Phi(\Gamma)$.

 The tensor product in the circling rules reflects the partition of the vertex set of the Feynman diagram
 $\Gamma$ and in the cutting rules it shows that the cutting propagator decomposes into the integration
 over incoming and outgoing external lines, while in the Connes--Kreimer Hopf algebra it disentangles
 overlapping divergences or reduces nested divergences. That is to say that the coproducts in the
 circling rules and in the cutting rules are compatible with the Connes--Kreimer Hopf
 algebra,
 \eq
 \Delta (S_R\star\Phi(\Gamma))=((S_R\star\Phi)\otimes(S_R\star\Phi))\Delta(\Gamma).
 \en
The term on the left hand side expands \eq
\Delta(S_R\star\Phi(\Gamma))=\Delta(\Phi(\Gamma))+\Delta(S_R(\Gamma))+\sum
\,S_R(\gamma)\Delta(\Phi(\Gamma/\gamma)) \en where the global
counter term $S_R(\Gamma)$ plays as a vertex and counter terms
$S_R(\gamma)$ for subdivergences act as coefficients. On the right
hand side, the unit $e$ or undivergent subdiagrams $\gamma$
satisfy $S_R\star\Phi(e)=e, S_R\star\Phi(\gamma)=\gamma$. As an
example, the renormalized Feynman diagram  $\Gamma_{\mbox{\tiny
Fig.6}}$ has a coproduct
\begin{center}
  \begin{picture}(320,40)(0,0)
 \Text(-5,30)[]{$\Delta ($}
 \Vertex(10,30){1.5} \Vertex(30,30){1.5}
 \Text(10,26)[rt]{$x$} \Text(30,26)[lt]{$y$}
 \Oval(20,30)(5,10)(0)  \Text(60,30)[]{$)$}
 \Text(40,30)[]{$+$} \Vertex(50,30){2.}
 \Text(70,30)[]{$=$} \Text(80,30)[]{$($}
 \Vertex(90,30){1.5} \Vertex(110,30){1.5}
 \Text(90,26)[rt]{$x$} \Text(110,26)[lt]{$y$}
 \Oval(100,30)(5,10)(0) \Text(120,30)[]{$+$}
 \Vertex(130,30){2.} \Text(140,30)[]{$)$}
 \Text(150,30)[]{$\otimes$} \Text(160,30)[]{$e$}
 \Text(170,30)[]{$+$}\Text(180,30)[]{$e$} \Text(190,30)[]{$\otimes$}
 \Text(200,30)[]{$($}
 \Vertex(210,30){1.5} \Vertex(230,30){1.5}
 \Text(210,26)[rt]{$x$} \Text(230,26)[lt]{$y$}
 \Oval(220,30)(5,10)(0) \Text(240,30)[]{$+$}
 \Vertex(250,30){2} \Text(260,30)[]{$)$}
  \Text(270,30)[]{$+$}
 \Vertex(290,30){1.5}
  \Line(280,40)(289.5,30)  \Line(280,20)(289.5,30) \Text(290,26)[tl]{$y$}
  \Text(300,30)[]{$\otimes$} \Vertex(310,30){1.5}
  \Text(310,26)[tr]{$x$} \Line(311.5,30)(320,20) \Line(311.5,30)(320,40)
  \end{picture}
  \vspace{-0.5cm}
  {\\ Figure 13. An example for the coproduct of a renormalized Feynman diagram.}
  \end{center}
For the third term on the right hand side, the integration over
the phase space of on-shell particles being finite ensures that
the renormalized cutting equation is well-defined.

\section{Concluding Remarks}

For a Feynman diagram $\Gamma$, there is a universal coproduct \eq
\label{universal}
 \Delta(\Gamma)=\Gamma\otimes\,e+\, e\otimes \Gamma+\sum\,\gamma\otimes\Gamma/\gamma
 \en
where $\gamma$ denotes the subdiagram of $\Gamma$ and the
summation is over all nontrivial subdiagrams. When $\gamma$ is
specified by a subset of the vertex set of $\Gamma$, the coproduct
represents the circling rules for the largest time equation and
the antipode denotes the conjugation diagram $\Gamma^\ast$. When
$\gamma$ denotes an admissible cut diagram, the coproduct leads to
the cutting rules for the cutting equation and the antipode is an
advanced function vanishing in retarded regions. When $\gamma$ is
required to be the disjoint union of divergent subdiagrams, it
represents the coproduct of the Connes--Kreimer Hopf algebra
\cite{kreimer1, kreimer2, kreimer3} and its twisted antipode is
the Zimmermann's forest formula \cite{zim1}.

The Hopf algebraic structures in proving the perturbative
unitarity have been considered in scalar field theories. They are
expected to be found in other field theories \cite{vanniu}. In
fermionic field theories, the same Hopf algebraic structures will
be obtained. But in gauge field theories, there are some subtle.
The unphysical degrees of freedom such as ghosts and time-like
component of gauge potential are involved in the cutting equation.
They have to be removed by applying the Ward identities or the
Slavnov--Taylor identities \cite{thooft2, vanniu, sterman}. The
meaningful coproduct has to be defined for the set of Feynman
diagrams. Therefore in the universal coproduct, $\Gamma$ may take
the summation of $\Gamma_i$, namely $\Gamma=\sum_i\,\Gamma_i$.

Besides the renormalizability and unitarity, physically
interesting quantum field theories have to satisfy the causality
principle. With the admissible cut diagrams, the dispersion
relation representing the causality takes the form of two-largest
time equation \cite{thooft2, vanniu}. The Hopf algebraic
structures similar to the case of the perturbative unitarity are
expected to be found. In addition, the dispersion equation can be
used to prove the power counting theorem or the locality of
counter terms. The Hopf algebraic structures may play the
important role since the Connes--Kreimer Hopf algebra solves the
locality of counter terms.

\vspace{1cm}


\begin{center}
\section*{Acknowledgments}
\end{center}

 We thank P. van Nieuwenhuizen for his helpful lectures on the cutting rules in
 the ``Mitteldeutsche Physik-Combo'' and thank J.P. Ma for helpful discussions.
 We thank D. Kreimer and X.Y. Li for helpful comments.

\appendix

\vspace{2cm}

\section{The axioms of the Hopf algebra}

Let $(H,+,m,\eta,\Delta,\eps,S;F)$ be a Hopf algebra over $F$. We
denote the set by $H$, the field by $ F$, the addition by $+$, the
product by $m$, the unit map by $\eta$, the coproduct by $\Delta$,
the counit by $\eps$, the antipode by $S$, the identity map by
$\Id$, and the tensor product by $\otimes$. The Hopf algebra has
to satisfy the following seven axioms: $(1)~~
m(m\otimes{\Id})=m({\Id}\otimes m), ~m: H\otimes H \to H,~
m(a\otimes b)=ab,~a, b\in H$;
$(2)~~m({\Id}\otimes\eta)={\Id}=m(\eta\otimes {\Id}), ~\eta: {
 F}\to H $  which denote the associative product $m$ and the
linear unit map $\eta$ in the algebra $(H,+,m,\eta; F)$ over $ F$
respectively;
$(3)~~(\Delta\otimes{\Id})\Delta=({\Id}\otimes\Delta)\Delta,
~\Delta: H \to H\otimes H,$
 $(4)~~({\Id}\otimes \epsilon)\Delta={\Id}=(\epsilon\otimes{\Id})\Delta,
 ~\eps: H\to { F} $ which denote the coassociative
coproduct $\Delta$ and the linear counit map $\eps$ in the
coalgebra $(H,+,\Delta,\eps; F)$ over $ F$ and where we have used
\sss{ F\otimes H}{H} or \sss{H\otimes  F}{H}; $
(5)~~\Delta(ab)=\Delta(a)\Delta(b), (6)~~
\eps(ab)=\eps(a)\eps(b),$ $\Delta(e)=e\otimes e,~\eps(e)=1, ~
a,b,e\in H $ which are the compatibility conditions between the
algebra and the coalgebra in the bialgebra
$(H,+,m,\eta,\Delta,\eps; F)$ over $ F$, claiming that the
coproduct $\Delta$ and the counit $\epsilon$ are homomorphisms of
the algebra $(H,+,m,\eta; F)$ over $ F$ with the unit $e$; $
 (7)~~m(S\otimes{\Id})\Delta=\eta\circ\epsilon=m({\Id}\otimes S)\Delta
$ which is the antipode axiom that can be used to define the
antipode. The character $\Phi$ is a nonzero linear functional over the
algebra and is a homomorphism satisfying
\sss{\Phi(ab)}{\Phi(a)\Phi(b)}. The convolution between two
characters $\Phi$ and $\Phi_c$ is defined by $\Phi\star
\Phi_c:=m(\Phi\otimes \Phi_c)\Delta$.

As an example, the coproduct for the matrix entry $A_{ij}$ of the
$n\times n$ matrix $A$ is defined as
$\Delta(A_{ij})=\sum_{k=1}^n\,A_{ik}\otimes\,A_{kj}$, while the
multiplication has $\sum_{k=1}^n\,A_{ik}\,A_{kj}=A_{ij}$. The
counit $\epsilon$ is defined by $\epsilon(A_{ij})=\delta_{ij}$ and
the antipode of $A_{ij}$ is its inverse $A^{-1}_{ij}$.  In our
case, the following type of coproduct \eq
\Delta(A_{ij})=A_{ij}\otimes e\,+\, e\otimes
A_{ij}\,+\,\sum_{k=1}^n\,A_{ik}\otimes\,A_{kj} \en is considered.

\end{document}